\documentclass[pra,showpacs,twocolumn]{revtex4-1}%
\usepackage{amsfonts}
\usepackage{amsmath}
\usepackage{amssymb,amscd}
\usepackage{psfrag}
\usepackage{graphicx}
\usepackage{braket}
\usepackage[dvips]{epsfig}
\usepackage{epsfig}
\usepackage{subfigure}
\usepackage{color}
\usepackage{amssymb}%
\providecommand{\U}[1]{\protect\rule{.1in}{.1in}}
\begin{document}
\title{Phase estimation for an SU(1,1) interferometer in the presence of phase
diffusion and photon losses}
\author{Xiu-Ling Hu$^{1}$}
\author{Dong Li$^{2,3}$}
\author{L. Q. Chen$^{1}$}
\email{lqchen@phy.ecnu.edu.cn}
\author{Keye Zhang$^{1}$}
\email{kyzhang@phy.ecnu.edu.cn}
\author{Weiping Zhang$^{4,5}$}
\author{Chun-Hua Yuan$^{1,5}$}
\email{chyuan@phy.ecnu.edu.cn}

\address{$^1$Quantum Institute for Light and Atoms, Department of Physics, East China
Normal University, Shanghai 200062, P. R. China}

\address{$^2$Microsystems and Terahertz Research Center, China Academy of Engineering Physics, Chengdu, Sichuan 610200, China}

\address{$^3$Institute of Electronic Engineering, China Academy of Engineering Physics, Mianyang, Sichuan 621999, China}

\address{$^4$Department of Physics, Shanghai Jiao Tong University, and Tsung-Dao Lee Institute, Shanghai 200240, China}

\address{$^5$Collaborative Innovation Center of Extreme Optics, Shanxi
University, Taiyuan, Shanxi 030006, P. R. China}

\begin{abstract}
We theoretically study the quantum Fisher information (QFI) of the SU(1,1)
interferometer with phase shifts in two arms taking account of realistic noise
effects. A generalized phase transform including the phase diffusion effect is
presented by the purification process. Based on this transform, the analytical
QFI and the bound to the quantum precision are derived when considering the
effects of phase diffusion and photon losses simultaneously. To beat the
standard quantum limit with the reduced precision of phase estimation due to
noisy, the upper bounds of decoherence coefficients as a function of total
mean photon number are given.

\end{abstract}
\date{\today }

\maketitle

\section{Introduction}

The estimation of the optical phase has been proposed and demonstrated using
different quantum strategies and interferometric setups
\cite{Helstrom67,Holevo82,Caves81,Braunstein94,Braunstein96,Lee02,Giovannetti06,Zwierz10,Giovannetti04,Giovannetti11,Toth,Degen}%
.\ In particular, interferometers can provide the most precise measurements,
for example, the gravitational waves can be observed by the advanced Laser
Interferometer Gravitational-Wave Observatory (LIGO)~\cite{Abbott}. The
Mach-Zehnder interferometer (MZI) and its variants have been used as a generic
model to realize precise measurement of phase. In 1986, an SU(1,1)
interferometer was introduced by Yurke \textit{et al.}~\cite{Yurke86}, where
two nonlinear beam splitters (NBSs) take the place of two linear beam
splitters (BSs) in the traditional MZI. It is called the SU(1,1)
interferometer because it is described by the SU(1,1) group, as opposed to the
SU(2) one for BSs. The phase sensitivity of an SU(1,1) interferometer scales
as $1/n$ instead of $1/\sqrt{n}$, where $n$ is the average number of photons
inside the interferometer. This enhanced sensitivity has attracted the
attention of many researchers to study this interferometer, which have been
constructed with several platforms, such as light
\cite{Plick,Jing11,Hudelist,Du18}, atoms
\cite{Gross,Peise,Gabbrielli,Linnemann16}, nonlinear crystals
\cite{Lemieux16,Manceau17}, and matter-light hybrid
\cite{ChenPRL15,Yama,Haine,Szigeti,Haine16,Yuan16,Barzanjeh}. Recently, a
simplified variation of the SU(1,1) interferometer-truncated SU(1,1)
interferometer has been presented, which can achieve the same phase
sensitivity \cite{Anderson1,Anderson2,Gupta}.

For an SU(1,1) interferometer, several measurement methods, such as intensity
detection \cite{Ou,Marino}, homodyne detection \cite{Ou,Li14} and parity
detection \cite{Li16}, have been studied to analyze phase estimation. Every
measurement method has its own superiority and these several detection methods
were compared \cite{Li16,Li18}. In general, it is difficult to optimize over
the detection methods to obtain the optimal estimation protocols. Fortunately,
the quantum Fisher information (QFI) is the intrinsic information in the
quantum state and is not related to the actual measurement procedure
\cite{Braunstein94,Braunstein96}. It characterizes the maximum amount of
information that can be extracted from quantum experiments about an unknown
parameter using the best (and ideal) measurement device. It establishes the
best precision that can be attained with a given quantum probe. We have
investigated the QFIs of the SU(1,1) interferometer with several different
input states for lossless case \cite{Li16,Gong17c}.

Because there are inevitable interactions with the surrounding environment, in
the presence of environment noise, the QFI and the corresponding measurement
precision will be reduced, which have been studied by many researchers
\cite{Dem09,Dem12,Escher12,Berry13, Chaves, Dur,
Kessler,Brivio,Alipour,Escher11,Genoni11,Genoni12,Sparaciari,Feng}. For
interferometers, there are three typical decoherence processes that should be
taken into account: (1) photon losses, which may happen at any stage of the
phase process and is modeled by the fictitious BS introduced in the
interferometer arms. The effect of photon losses on the interferometry has
been studied \cite{Dem09,Dem12,Escher11}, where this decoherence process can
be described by a set of Kraus operators. (2) phase diffusion, which
represents the effect of fluctuation of the estimated phase delay. Recently,
Escher \emph{et al.} \cite{Escher12} presented a variational approach to show
an analytical bound based on purification techniques, which has an explicit
dependence on the mathematical description of the noise. In some physical
processes, such as path length fluctuations of a stabilized interferometer,
thermal fluctuations of an optical fibre, phase and phase diffusion may vary
in time \cite{Trapani}. The weak measurements and joint estimation were used
to deal the phase and phase diffusion simultaneously for quantum metrology
\cite{Vidrighin,Altorio}. Recently, using this variational approach, Zwierz
and Wiseman extended the phase diffusion noise from linear to nonlinear and
gave the precision bound for a second-order phase-diffusion noise
\cite{Zwierz14}. (3) detection losses \cite{Ou,Sparaciari}. The influence of
the detection loss on the phase sensitivity is less than the photon loss
inside the interferometer. As for SU(1,1) interferometers, the influence of
detection loss can be suppressed by using unbalanced gains in the two
nonlinear processes \cite{Manceau17, Giese}.

In addition to the detection losses, the photon losses and phase diffusion
need to be investigated. Although, the photon losses effect on phase
estimation in an SU(1,1) interferometer have been studied
\cite{Ou,Marino,Gong17}. However, a more accurate and efficient scheme is
required to estimate the phase shift considering the effects of phase
diffusion and photon losses simultaneously. In this paper, we derive the QFI
of the SU(1,1) interferometer in the presence of phase diffusion and photon
losses simultaneously based on the generalized phase transform. Then the phase
estimation precision as a function of the decoherence coefficients is
discussed and the upper bound of decoherence coefficients to beat the standard
quantum limit is given.

Our article is organized as follows. In Sec. II, we briefly describe the
generalized decoherence model. In Sec. III, the form of generalized phase
transform is given based on the purification process, based on which the Kraus
operators including the photon losses and phase diffusion coefficients are
derived. In Sec. IV, the QFI of the SU(1,1) interferometer in the presence of
losses and phase diffusion simultaneously is obtained and discussed. Finally,
we conclude with a summary of our results.

\begin{figure}[ptbh]
\centerline{\includegraphics[scale=0.7,angle=0]{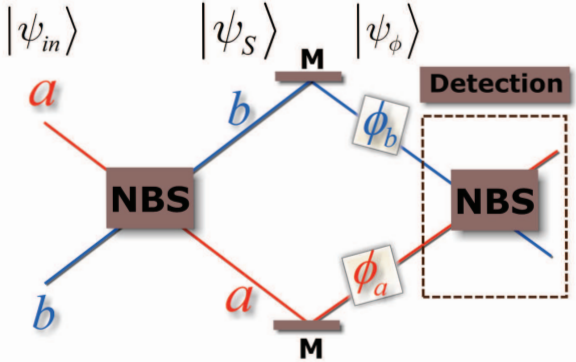}}\caption{ Schematic
diagram of an SU(1,1) interferometer. $|\psi_{in}\rangle$, $|\psi_{S}\rangle$
and $|\psi_{\phi}\rangle$ are the different states of the evolution process
without decoherence. $\phi_{a}$ and $\phi_{b}$ are the phase shifts arms $a$
and $b$, respectively.}%
\label{fig1}%
\end{figure}

\section{Decoherence model of SU(1,1) interferometers}

Decoherence is a consequence of the uncontrolled interactions of a quantum
system with the environment. Effects of decoherence inside an interferometer
should be taken into account. Recently, Escher \textit{et al}. \cite{Escher11}
developed a general formalism in the presence of losses. Here, we extend this
model to derive the QFI in the presence of losses and phase diffusion simultaneously.

Given an initial pure state $\hat{\rho}_{0S}=|\psi_{S}\rangle\langle\psi_{S}|$
of a system $S$, in the presence of losses and phase diffusion the evolution
of system is non-unitary. However, in a bigger space the final enlarged state
is given by%
\begin{equation}
\hat{\rho}_{S,E,E^{\prime}}=\hat{U}_{S,E,E^{\prime}}(\eta,\beta,\phi)\hat
{\rho}_{0S}\otimes\hat{\rho}_{0E}\otimes\hat{\rho}_{0E^{\prime}}\hat
{U}_{S,E,E^{\prime}}^{\dag},
\end{equation}
where $\hat{U}_{S,E,E^{\prime}}(\eta,\beta,\phi)$ is the corresponding unitary
operator of the enlarged state ($S+E+E^{\prime}$). $\hat{\rho}_{0E}%
=|0_{E}\rangle\langle0_{E}|$, $|0_{E}\rangle$\ is the initial state of the
phase diffusion model, and $\beta$\ is the phase diffusion coefficient
($\beta=0$, corresponding to no diffusion; $\beta=\infty$, corresponding to
maximum diffusion). $\hat{\rho}_{0E^{\prime}}=|0_{E^{\prime}}\rangle
\langle0_{E^{\prime}}|$, $|0_{E^{\prime}}\rangle$\ is the initial state of the
photon loss environment. $\eta$ quantifies the photon losses (from $\eta=1$,
lossless case, to $\eta=0$, complete absorption).

When the environment $E^{\prime}$ is not monitored, one may take the partial
trace with respect to $E^{\prime}$,%
\begin{align}
\hat{\rho}_{S,E}  &  =\mathrm{Tr}_{E^{\prime}}\{\hat{U}_{S,E,E^{\prime}}%
(\eta,\beta,\phi)[\hat{\rho}_{0S}\otimes\hat{\rho}_{0E}\otimes\hat{\rho
}_{0E^{\prime}}]\hat{U}_{S,E,E^{\prime}}^{\dag}\}\nonumber\\
&  =\sum_{l}\langle l_{E^{\prime}}|\hat{U}_{S,E,E^{\prime}}(\eta,\beta
,\phi)[\hat{\rho}_{0S}\hat{\rho}_{0E}\hat{\rho}_{0E^{\prime}}]\hat
{U}_{S,E,E^{\prime}}^{\dag}|l_{E^{\prime}}\rangle\nonumber\\
&  =\sum_{l}\hat{\Pi}_{l}(\eta,\beta,\phi)\hat{\rho}_{0S}\hat{\rho}_{0E}%
\hat{\Pi}_{l}^{\dag}(\eta,\beta,\phi),
\end{align}
where $|l_{E^{\prime}}\rangle$\ are the orthogonal states of the environment
$E^{\prime}$, and $\hat{\Pi}_{l}(\eta,\beta,\phi)=\langle l_{E^{\prime}}%
|\hat{U}_{S,E,E^{\prime}}(\eta,\beta,\phi)|0_{E^{\prime}}\rangle$ are Kraus
operators which not only describe the photon losses process, but also include
the phase evolution. Without considering photon losses and only considering
the phase diffusion, based on the purified process the enlarged
system-environment ($S+E$) state is written as%
\begin{equation}
\left\vert \Phi_{S,E}(\phi)\right\rangle =\tilde{U}(\phi,\beta)\left\vert
\psi_{S}\right\rangle \left\vert 0_{E}\right\rangle ,
\end{equation}
where $\tilde{U}(\phi,\beta)$ is the generalized phase transformation
including the phase diffusion coefficient $\beta$ compared to the phase
transform $\hat{U}(\phi)$ of the ideal interferometer. Usually, $\hat{\Pi}%
_{l}(\eta,\phi,\beta)$ may be written as $\hat{\Lambda}_{l}(\eta)\tilde
{U}(\phi,\beta)$ or $\tilde{U}(\phi,\beta)\hat{\Lambda}_{l}(\eta)$ where
$\hat{\Lambda}_{l}(\eta)$\ are Kraus operators without considering the phase
shift. In realistic systems the photon losses are distributed throughout the
arms of interferometer, not an event concentrated either before or after the
phase $\tilde{U}(\phi,\beta)$ is displaced. Then $\hat{\Lambda}_{l}%
(\eta)\tilde{U}(\phi,\beta)$ or $\tilde{U}(\phi,\beta)\hat{\Lambda}_{l}(\eta)$
are just a matter of convenience. An optimized Kraus operator $\hat{\Pi}%
_{l}(\eta,\beta,\phi)$ can be obtained by making the QFI is minimal, which
will be given in the next section.

For the enlarged system-environment ($S+E+E^{\prime}$) state $|\Psi
_{SEE^{\prime}}\rangle=\hat{U}_{SEE^{\prime}}(\eta,\beta,\phi)|\psi_{S}%
\rangle|0_{E^{\prime}}\rangle|0_{E}\rangle$, the QFI $\mathcal{C}_{Q}$ is
given by \cite{PezzBook}%
\begin{align}
\mathcal{C}_{Q}  &  =4\langle\Delta\hat{H}_{S,E,E^{\prime}}^{2}\rangle
=4[\langle0_{E^{\prime}}|\langle0_{E}|\langle\psi_{S}|\hat{H}_{S,E,E^{\prime}%
}^{2}|\psi_{S}\rangle|0_{E}\rangle|0_{E^{\prime}}\rangle\nonumber\\
&  -\langle0_{E^{\prime}}|\langle0_{E}|\langle\psi_{S}|\hat{H}_{S,E,E^{\prime
}}|\psi_{S}\rangle|0_{E}\rangle|0_{E^{\prime}}\rangle^{2}], \label{eq4}%
\end{align}
where $\hat{H}_{S,E,E^{\prime}}=i(d\hat{U}_{S,E,E^{\prime}}^{\dag}(\eta
,\beta,\phi)/d\phi)\hat{U}_{S,E,E^{\prime}}(\eta,\beta,\phi)$. The expression
$\langle\Delta\hat{H}_{S,E,E^{\prime}}^{2}\rangle$ may be rewritten as%
\begin{align}
\langle\Delta\hat{H}_{S,E,E^{\prime}}^{2}\rangle &  =\left(  \frac
{d\langle\Psi_{SEE^{\prime}}|}{dx}\right)  \left(  \frac{d|\Psi_{SEE^{\prime}%
}\rangle}{dx}\right) \nonumber\\
&  -\left\vert \left(  \frac{d\langle\Psi_{SEE^{\prime}}|}{dx}\right)
|\Psi_{SEE^{\prime}}\rangle\right\vert ^{2}.
\end{align}
Therefore, in terms of the initial state of the system $|\psi_{S}\rangle$, the
initial state of the phase diffusion model $|0_{E}\rangle$, and the Kraus
operators $\hat{\Pi}_{l}(\eta,\beta,\phi)$, the enlarged state can be written
as $|\Psi_{SEE^{\prime}}\rangle=\sum_{l}\hat{\Pi}_{l}(\eta,\beta,\phi
)|\psi\rangle_{S}\left\vert 0_{E}\right\rangle |l_{E^{\prime}}\rangle$, then
the value of $\mathcal{C}_{Q}$ is given by \cite{Escher11}
\begin{align}
\mathcal{C}_{Q}  &  =4\langle\Delta\hat{H}_{S,E,E^{\prime}}^{2}\rangle
=4[\langle0_{E}|\langle\psi_{S}|\hat{H}_{1}|\psi_{S}\rangle|0_{E}%
\rangle\nonumber\\
&  -\left\vert \langle0_{E}|\langle\psi_{S}|\hat{H}_{2}|\psi_{S}\rangle
|0_{E}\rangle\right\vert ^{2}], \label{CQ}%
\end{align}
where
\begin{align}
\hat{H}_{1}  &  =\sum_{l}\frac{d\hat{\Pi}_{l}^{\dag}(\eta,\beta,\phi)}{d\phi
}\frac{d\hat{\Pi}_{l}(\eta,\beta,\phi)}{d\phi},\text{\ }\nonumber\\
\hat{H}_{2}  &  =i\sum_{l}\frac{d\hat{\Pi}_{l}^{\dag}(\eta,\beta,\phi)}{d\phi
}\hat{\Pi}_{l}(\eta,\beta,\phi).
\end{align}
Because the additional freedom supplied by the environment ($E+E^{\prime}$)
should increase the QFI. Therefore, the QFI $\mathcal{F}_{Q}$ only describing
the system\ should be smaller or equal to the $\mathcal{C}_{Q}$ obtained on
the system plus environment ($E+E^{\prime}$). It is possible to determine the
value of QFI $\mathcal{F}_{Q}[\hat{\rho}_{S}]$ by two steps minimizing the
upper bound $\mathcal{C}_{Q}[\hat{\rho}_{S,E,E^{\prime}}]$. One is by
optimizing over all possible measurements $\hat{\Pi}_{l}$, the other is by the
minimum all possible purifications $\left\vert \Phi_{S,E}(\phi)\right\rangle
$. The relation between $\mathcal{F}_{Q}[\hat{\rho}_{S}]$\ and $\mathcal{C}%
_{Q}[\hat{\rho}_{S,E,E^{\prime}}]$ is written as
\begin{equation}
\mathcal{F}_{Q}=\underset{\{\hat{\Pi}_{l}(\eta,\beta,\phi),\text{ }\left\vert
\Phi_{S,E}(\phi)\right\rangle \}}{\min}\mathcal{C}_{Q}\left[  \hat{\rho}%
_{0S}\hat{\rho}_{0E},\hat{\Pi}_{l}(\eta,\beta,\phi)\right]  .
\end{equation}

\begin{figure}[tbh]
\centerline{\includegraphics[scale=0.6,angle=0]{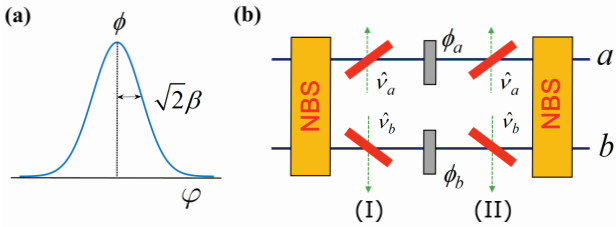}}\caption{ Lossy
interferometer model. (a) The phase diffusion corresponds to the application
of a random, Gaussian-distributed phase shift $\varphi$ with variance
$2\beta^{2}$\ and the mean equal to the estimated parameter $\phi$. (b) The
losses in the interferometer are modeled by adding fictitious beam splitters
(I) before and (II) after the phase shifts $\phi_{1}$ and $\phi_{2}$. $\hat
{v}_{a}$ and $\hat{v}_{b}$ represent the vacuum. }%
\label{fig2}%
\end{figure}

Next, we will describe the unitary phase transform $U(\phi)$, the generalized
phase transform $\tilde{U}(\phi,\beta)$, and the optimized Kraus operators
$\hat{\Pi}_{l}(\eta,\beta,\phi)$ including phase diffusion coefficients and
photon loss coefficients.

\subsection{Phase transformation $\hat{U}(\phi)$}

An SU(1,1) interferometer is shown in Fig.~\ref{fig1}, and the annihilation
operators of the two modes $a$, $b$ are denoted as $\hat{a}$, $\hat{b}$,
respectively. After the first NBS, the two beams sustain phase shifts, i.e.,
mode $a$ undergoes a phase shift of $\phi_{a}$\ and mode $b$ undergoes a phase
shift of $\phi_{b}$. In an ideal interferometer, the transformation of the
incoming state vector $\left\vert \psi_{S}\right\rangle $ in the
Schr\"{o}dinger picture is given as following
\begin{equation}
|\psi_{\phi}\rangle=\hat{U}(\phi)\left\vert \psi_{S}\right\rangle ,
\label{phase shift}%
\end{equation}
and%
\begin{align}
\hat{U}(\phi)  &  =\exp\left(  i\phi_{a}\hat{n}_{a}\right)  \exp(i\phi_{b}%
\hat{n}_{b})\nonumber\\
&  =\exp(i\phi\hat{K}_{z})\exp[i(\phi_{a}-\phi_{b})\hat{J}_{z}-i\phi/2],
\label{Uphi}%
\end{align}
where $\phi=\phi_{a}+\phi_{b}$, $\hat{K}_{z}=(\hat{n}_{a}+\hat{n}_{b}+1)/2$,
$\hat{J}_{z}=(\hat{n}_{a}-\hat{n}_{b})/2$, $\hat{n}_{a}=\hat{a}^{\dagger}%
\hat{a}$, and $\hat{n}_{b}=\hat{b}^{\dagger}\hat{b}$. The operator
$e^{i(\phi_{a}-\phi_{b})\hat{J}_{z}}$ gives rise to phase factors which do not
contribute to the expectation values of number operators.

\subsection{Generalized phase transformation $\tilde{U}(\phi,\beta)$}

In an SU(1,1) interferometer, it is vital to take into account the unavoidable
influence of noise on the ultimate precision limits. The collective dephasing
or the phase noise represents the effect of fluctuation of the estimated phase
shift $\phi$, which is also termed as phase diffusion. Now, we derive the
generalized $\tilde{U}(\phi,\beta)$\ including the phase diffusion.

In the Markovian limit, the phase diffusion environment is static, and the
diffusion coefficient is a time-independent variance. When the noisy
environments cannot be described in terms of a Markovian master equation, the
time-dependent variance and the dynamical properties of the environment should
be considered \cite{Trapani}. The phase diffusion of state can be modeled by
the effects of the radiation pressure on the interferometer mirror, where the
light is being reflected from a mirror which position fluctuations are
randomly changing the effective optical length \cite{Escher11}. Assuming the
measuring time is much less than the vibration period of the mirror, the
evolution between the light field and the mirror is taken as
\begin{equation}
\exp[i(\sqrt{2}\beta)\hat{n}(\hat{b}_{E}+\hat{b}_{E}^{\dag})]=\exp[i2\beta
\hat{n}\hat{x}_{E}],
\end{equation}
where $\hat{n}$ is the number operator of light field, $\hat{x}_{E}$
($=(\hat{b}_{E}+\hat{b}_{E}^{\dag})/\sqrt{2}$) and $\hat{b}_{E}$ are the
position and annihilate operators of mirror. Assuming the mirrors as the
environments in the ground state of a quantum oscillator $\left\vert 0_{E_{a}%
}\right\rangle $ and $\left\vert 0_{E_{b}}\right\rangle $ before interaction
with the light beam, the final state of the combined system ($S+E$) of the
probe and the mirror is given by%
\begin{align}
\left\vert \Phi_{S,E}(\phi)\right\rangle  &  =\hat{U}_{S,E}(\phi,\beta
_{a},\beta_{b})|\psi_{S}\rangle|0_{E}\rangle=e^{i\phi_{a}\hat{n}_{a}}%
e^{i\phi_{b}\hat{n}_{b}}\nonumber\\
&  \times e^{i(2\beta_{a})\hat{n}_{a}\hat{x}_{E_{a}}}e^{i(2\beta_{b})\hat
{n}_{b}\hat{x}_{E_{b}}}\left\vert \psi_{S}\right\rangle \left\vert 0_{E_{a}%
}\right\rangle \left\vert 0_{E_{b}}\right\rangle \nonumber\\
&  =e^{i\phi(\hat{K}_{z}-\frac{1}{2})}e^{i(\phi_{a}-\phi_{b})\hat{J}_{z}%
}e^{i(2\beta_{a})\hat{n}_{a}\hat{x}_{E_{a}}}\nonumber\\
\times &  e^{i(2\beta_{b})\hat{n}_{b}\hat{x}_{E_{b}}}\left\vert \psi
_{S}\right\rangle \left\vert 0_{E_{a}}\right\rangle \left\vert 0_{E_{b}%
}\right\rangle ,\label{Pstate}%
\end{align}
where $\hat{U}_{S,E}(\phi,\beta_{a},\beta_{b})$ is the corresponding unitary
operator acting on the enlarged state. $\beta_{a}$ and $\beta_{b}$\ are
diffusion coefficients in arms $a$ and $b$, respectively. This particular
purification also leads to a trivial noiseless upper bound on the QFI as in
Ref~\cite{Escher12}. However, the purified unitary evolution is integrally
written as%
\begin{equation}
|\Psi_{S,E}\rangle=\hat{u}_{E}(x)|\Phi_{S,E}\rangle=\hat{u}_{E}(x)\hat
{U}_{S,E}(x)\left\vert \psi_{S}\right\rangle \left\vert 0_{E_{a}}\right\rangle
\left\vert 0_{E_{b}}\right\rangle ,
\end{equation}
where $\hat{u}_{E}$ is a unitary operator acting only on the $E$. To the same
evolution of system $S$, many evolution of systems of $S+E$ lead to possibly
different values of the QFI. Therefore, a stronger upper bound\ can be
obtained by minimizing $\mathcal{C}_{Q}[|\Psi_{S,E}\rangle]$ over all
purifications of the system $S$. Inserting the evolution operator $\hat{u}%
_{E}(x)\hat{U}_{S,E}(x)$ into Eq.~(\ref{eq4}) without considering the
enviroment $E^{\prime}$, we obtain $\hat{H}_{S,E}$ and the corresponding QFI
$\mathcal{C}_{Q}=4\langle\Delta\hat{H}_{S,E}^{2}\rangle$. As $\langle
\Delta\hat{H}_{S,E}^{2}\rangle$\ only depends on $\hat{u}_{E}(x)$\ through
$\hat{h}_{E}(x)$, which is defined as $\hat{h}_{E}(x)=i(d\hat{u}_{E}^{\dag
}/dx)\hat{u}_{E}$ \cite{Escher12}.

Next, we follow the method of Escher \textit{et al}. \cite{Escher12} to derive
an optimal $\hat{h}_{E}^{(\mathrm{opt})}$, from which we can obtain a tighter
upper bound to the QFI of the system $S$, and then the corresponding
generalized phase transform $\tilde{U}(\phi,\beta)$. The optimal Hermitian
operator $\hat{h}_{E}^{(\mathrm{opt})}$ is given by%
\begin{equation}
\frac{1}{2}\left(  \hat{h}_{E}^{(\mathrm{opt})}\hat{\rho}_{E}+\hat{\rho}%
_{E}\hat{h}_{E}^{(\mathrm{opt})}\right)  =\mathrm{Tr}_{S}\{D\left[  \hat{\rho
}_{S,E}\right]  \}, \label{Key}%
\end{equation}
where $\hat{\rho}_{E}=\mathrm{Tr}_{S}[\left\vert \Phi_{S,E}\right\rangle
\left\langle \Phi_{S,E}\right\vert ]=\mathrm{Tr}_{S}[\hat{\rho}_{S,E}]$ is the
reduced density matrix in the $E$ space, and $D\left[  \hat{\rho}%
_{S,E}\right]  $ is defined as%
\begin{equation}
D\left[  \hat{\rho}_{S,E}\right]  \equiv\frac{i}{2}\left[  \frac{d\left\vert
\Phi_{S,E}\right\rangle }{dx}\left\langle \Phi_{S,E}\right\vert -\left\vert
\Phi_{S,E}\right\rangle \frac{d\left\langle \Phi_{S,E}\right\vert }%
{dx}\right]  .
\end{equation}
From Eq. (\ref{Pstate}), the reduced density matrix of the mirror associated
with the purification $\left\vert \Phi_{S,E}(\phi)\right\rangle $ is
\begin{align}
\hat{\rho}_{E}  &  =\sum_{n_{a}}\sum_{n_{b}}\left\vert \rho_{n_{a}n_{b}%
}\right\vert ^{2}\left\vert i\sqrt{2}\beta_{a}n_{a}\right\rangle _{E_{a}%
}\left\langle i\sqrt{2}\beta_{a}n_{a}\right\vert \nonumber\\
&  \times\left\vert i\sqrt{2}\beta_{b}n_{b}\right\rangle _{E_{b}}\left\langle
i\sqrt{2}\beta_{b}n_{b}\right\vert ,
\end{align}
and%
\begin{align}
\mathrm{Tr}_{S}\{D\left[  \hat{\rho}_{S,E}\right]  \}  &  =-\frac{1}{2}%
\sum_{n_{a}}\sum_{n_{b}}[(\frac{-i\hat{b}_{E_{a}}}{2\sqrt{2}\beta_{a}}%
+\frac{-i\hat{b}_{E_{b}}}{2\sqrt{2}\beta_{b}})\hat{\rho}_{E}\nonumber\\
&  +\hat{\rho}_{E}(\frac{i\hat{b}_{E_{a}}^{\dag}}{2\sqrt{2}\beta_{a}}%
+\frac{i\hat{b}_{E_{b}}^{\dag}}{2\sqrt{2}\beta_{b}})],
\end{align}
where $\hat{b}_{E_{i}}=(\hat{x}_{E_{i}}+i\hat{p}_{E_{i}})/\sqrt{2}$ with
$\hat{p}_{E_{i}}$ being the dimensionless momentum operator of mirror
($i=a,b$). Therefore, the Eq.~(\ref{Key}) can be rewritten as%
\begin{align}
&  \frac{1}{2}\left(  \hat{h}_{E}^{(\mathrm{opt})}\hat{\rho}_{E}+\hat{\rho
}_{E}\hat{h}_{E}^{(\mathrm{opt})}\right)  =-\frac{1}{2}\sum_{n_{a}}\sum
_{n_{b}}[(\frac{\hat{p}_{E_{a}}}{4\beta_{a}}+\frac{\hat{p}_{E_{b}}}{4\beta
_{b}})\hat{\rho}_{E}\nonumber\\
&  +\hat{\rho}_{E}(\frac{\hat{p}_{E_{a}}}{4\beta_{a}}+\frac{\hat{p}_{E_{b}}%
}{4\beta_{b}})-i(\frac{\hat{x}_{E_{a}}}{4\beta_{a}}+\frac{\hat{x}_{E_{b}}%
}{4\beta_{b}})\hat{\rho}_{E}\nonumber\\
&  +i\hat{\rho}_{E}(\frac{\hat{x}_{E_{a}}}{4\beta_{a}}+\frac{\hat{x}_{E_{b}}%
}{4\beta_{b}})]. \label{Key2}%
\end{align}
To get a tighter upper bound to the QFI of the system $S$, according to the
Eq.~(\ref{Key2}) we can guess $\hat{h}_{E}^{(\mathrm{opt})}=\lambda(\hat
{p}_{E_{a}}/4\beta_{a}+\hat{p}_{E_{b}}/4\beta_{b})$, and we can also obtain
the optimal $\hat{u}_{E}(x)=\exp[i\phi\lambda(\hat{p}_{E_{a}}/4\beta_{a}%
+\hat{p}_{E_{b}}/4\beta_{b})]$ with $\lambda$ being a variational parameter
\cite{Escher11}. The new purification process produces the following enlarged
system-environment ($S+E$) state:
\begin{equation}
\left\vert \Phi_{S,E}(\phi)\right\rangle =\tilde{U}(\phi,\beta_{a},\beta
_{b})\left\vert \psi_{S}\right\rangle \left\vert 0_{E_{a}}\right\rangle
\left\vert 0_{E_{b}}\right\rangle ,
\end{equation}
where the generalized phase transform $\tilde{U}(\phi,\beta_{a},\beta_{b})$ is
given by%
\begin{align}
\tilde{U}(\phi,\beta_{a},\beta_{b})  &  =e^{i\phi\lbrack\lambda(\hat{p}%
_{E_{a}}/4\beta_{a}+\hat{p}_{E_{b}}/4\beta_{b})+\hat{K}_{z}-\frac{1}{2}%
]}e^{i(\phi_{a}-\phi_{b})\hat{J}_{z}}\nonumber\\
&  \times e^{i(2\beta_{a})\hat{n}_{a}\hat{x}_{E_{a}}}e^{i(2\beta_{b})\hat
{n}_{b}\hat{x}_{E_{b}}}. \label{G-Uphi}%
\end{align}

\subsection{Kraus operators with noisy}

Here, we derive the Kraus operators $\hat{\Pi}_{l}(\eta,\beta,\phi)$ including
phase diffusion coefficients and photon loss coefficients based on the
generalized phase transform $\tilde{U}(\phi,\beta_{a},\beta_{b})$.

As shown in Fig.~\ref{fig2}(b), beam splitters are used as a model for the
photon-loss mechanism \cite{Dem09}. After the photon passing through the
virtual beam splitter, the evolution of the state can be described by a Kraus
operator $\hat{\Lambda}_{l_{a}l_{b}}(\eta_{a},\eta_{b})=\sqrt{\frac
{(1-\eta_{a})^{l_{a}}(1-\eta_{b})^{l_{b}}}{l_{a}!l_{b}!}}\eta_{a}^{\hat{n}%
_{a}/2}\eta_{b}^{\hat{n}_{b}/2}\hat{a}^{l_{a}}\hat{b}^{l_{b}}$, where
$\eta_{i}$ quantifies the photon losses of arm $i$ ($\eta_{i}=1$, lossless
case, $\eta_{i}=0$, complete absorption, $i=a,b$). Corresponding to the photon
losses before or after the phase shifts, the Kraus operators including the
generalized phase factor $\tilde{U}(\phi,\beta_{a},\beta_{b})$ are written as
\begin{align}
\hat{\Pi}_{l_{a}l_{b}}(\eta_{a},\eta_{b},\beta_{a},\beta_{b},\phi)  &
=\sqrt{\frac{(1-\eta_{a})^{l_{a}}(1-\eta_{b})^{l_{b}}}{l_{a}!l_{b}!}%
}\nonumber\\
&  \times\tilde{U}(\phi,\beta_{a},\beta_{b})\eta_{a}^{\frac{\hat{n}_{a}}{2}%
}\eta_{b}^{\frac{\hat{n}_{b}}{2}}\hat{a}^{l_{a}}\hat{b}^{l_{b}},
\end{align}
or
\begin{align}
\hat{\Pi}_{l_{a}l_{b}}(\eta_{a},\eta_{b},\beta_{a},\beta_{b},\phi)  &
=\sqrt{\frac{(1-\eta_{a})^{l_{a}}(1-\eta_{b})^{l_{b}}}{l_{a}!l_{b}!}%
}\nonumber\\
&  \times\eta_{a}^{\frac{\hat{n}_{a}}{2}}\eta_{b}^{\frac{\hat{n}_{b}}{2}}%
\hat{a}^{l_{a}}\hat{b}^{l_{b}}\tilde{U}(\phi,\beta_{a},\beta_{b}).
\end{align}

\begin{figure}[tbh]
\centerline{\includegraphics[scale=0.42,angle=0]{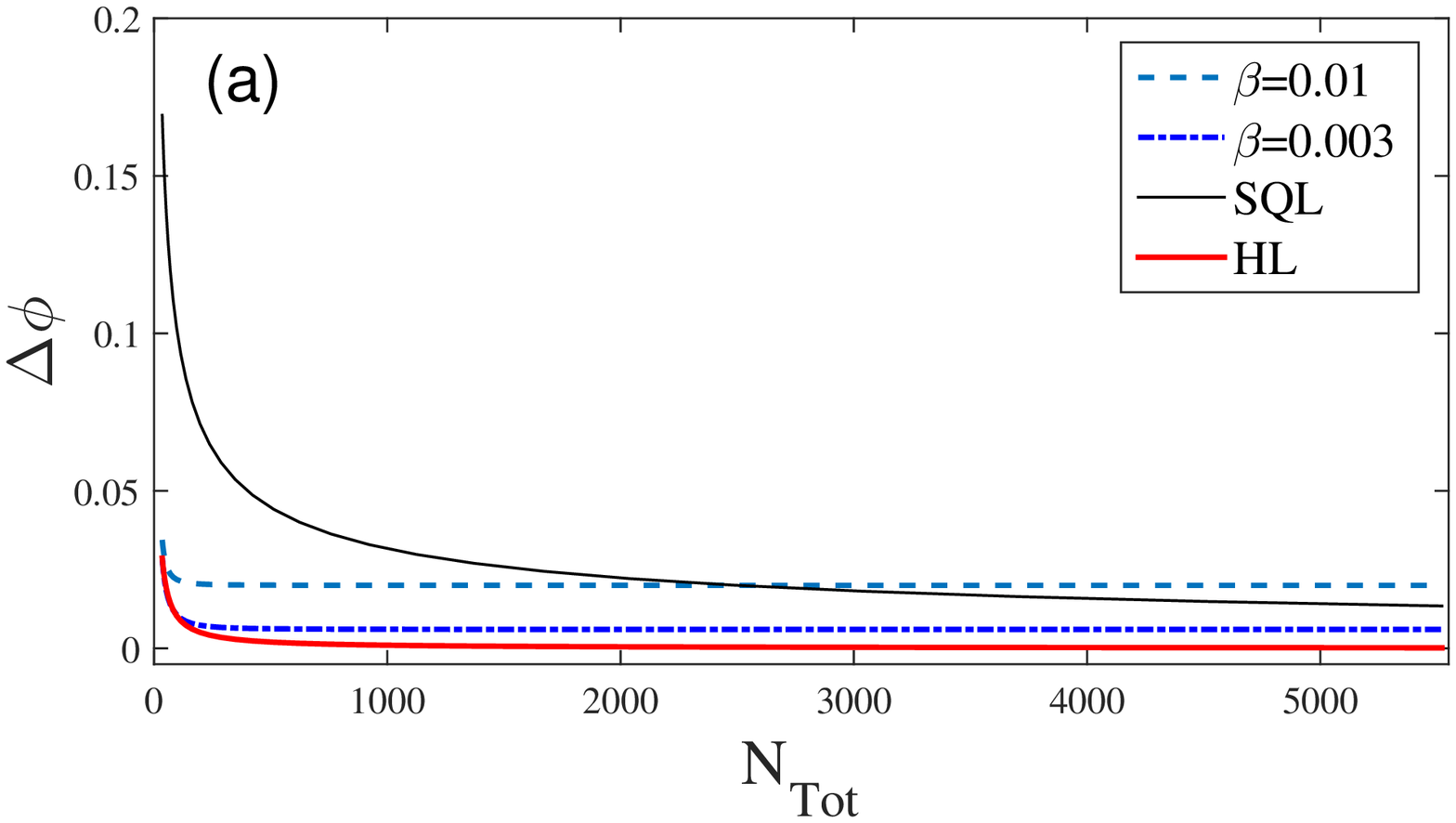}}
\centerline{\includegraphics[scale=0.42,angle=0]{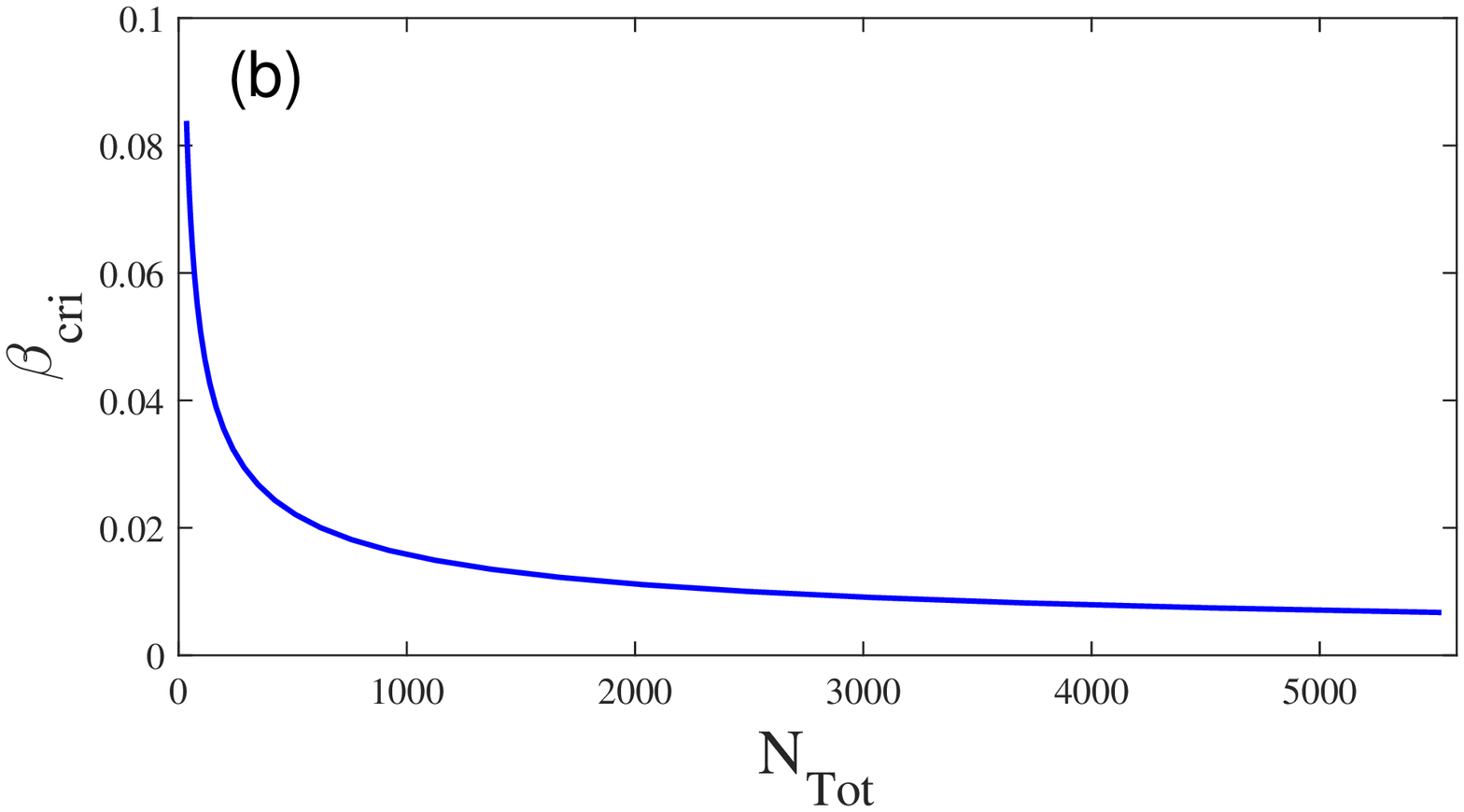}} \caption{(a) The
phase sensitivity $\Delta\phi$ versus $N_{\mathrm{Tot}}$ with $\beta=0.01$
(dashed line) and $\beta=0.003$ (dotted-dashed line), $g=2$. (b) The critical
$\beta_{\mathrm{cri}}$ as a function of the total mean photon number inside
the interferometer $N_{\mathrm{Tot}}$ where $g=2$, and $\beta_{a}$=$\beta_{b}%
$. Given a $N_{\mathrm{Tot}}$, the phase sensitivity $\Delta\phi$ beat the SQL
when $\beta<\beta_{\mathrm{cri}}$.}%
\label{fig3}%
\end{figure}

By introducing two parameters $\gamma_{a}$ and $\gamma_{b}$, we define a
family of Kraus operators, which can be used to minimize the value of
$\mathcal{C}_{Q}$. Then a possible set of Kraus operators describing the phase
diffusion and photon losses is given by%
\begin{align}
&  \hat{\Pi}_{l_{a}l_{b}}(\eta_{a},\eta_{b},\beta_{a},\beta_{b},\phi
)=\sqrt{\frac{(1-\eta_{a})^{l_{a}}(1-\eta_{b})^{l_{b}}}{l_{a}!l_{b}!}%
}\nonumber\\
&  \times e^{i(\phi_{1}-\phi_{2})(\hat{J}_{z}-\frac{\gamma_{a}l_{a}}{2}%
+\frac{\gamma_{b}l_{b}}{2})}\nonumber\\
&  \times e^{i\phi\lbrack(\frac{\hat{n}_{a}-\gamma_{a}l_{a}}{2}+\lambda
\frac{\hat{p}_{E_{a}}}{4\beta a})+(\frac{\hat{n}_{b}-\gamma_{b}l_{b}}%
{2}+\lambda\frac{\hat{p}_{E_{b}}}{4\beta_{b}})]}\nonumber\\
&  \times e^{i[(\beta_{a}\hat{x}_{E_{a}}-\beta_{b}\hat{x}_{E_{b}})(\hat{n}%
_{a}-\gamma_{a}l_{a}-\hat{n}_{b}+\gamma_{b}l_{b})]}\nonumber\\
&  \times e^{i[(\beta_{a}\hat{x}_{E_{a}}+\beta_{b}\hat{x}_{E_{b}})(\hat{n}%
_{a}-\gamma_{a}l_{a}+\hat{n}_{b}-\gamma_{b}l_{b})]}\eta_{a}^{\frac{\hat{n}%
_{a}}{2}}\eta_{b}^{\frac{\hat{n}_{b}}{2}}\hat{a}^{l_{a}}\hat{b}^{l_{b}},
\label{Kraus}%
\end{align}
where $\gamma_{i}$ describes\ the photon loss before ($\gamma_{i}=0$) and
after ($\gamma_{i}=-1$) the phase shifts of arm $i$ ($i=a,b$).

\section{QFI of SU(1,1) interferometers}

In this section, we use the above Kraus operators to obtain the QFI in the
presence of photon losses and phase diffusion simultaneously.

Without decoherence, the definition of QFI for pure states is simplified to
\cite{PezzBook}%
\begin{equation}
\mathcal{F}_{Q}=4(\left\langle \psi_{\phi}^{\prime}\right\vert \psi_{\phi
}^{\prime}\rangle-\left\vert \left\langle \psi_{\phi}^{\prime}\right\vert
\psi_{\phi}\rangle\right\vert ^{2}),\text{ }\left\vert \psi_{\phi}^{\prime
}\right\rangle =\frac{d\left\vert \psi_{\phi}\right\rangle }{d\phi}.
\end{equation}
The QFI of an SU(1,1) interferometer can be worked out $\mathcal{F}%
_{Q}=4\langle\Delta^{2}\hat{K}_{z}\rangle$, where $\langle\Delta^{2}\hat
{K}_{z}\rangle=\left\langle \psi_{S}\right\vert \hat{K}_{z}^{2}|\psi
_{S}\rangle-\left\vert \left\langle \psi_{S}\right\vert \hat{K}_{z}|\psi
_{S}\rangle\right\vert ^{2}$. Further simplification, the QFI $\mathcal{F}%
_{Q}$ can be written as \cite{Gong17}
\begin{equation}
\mathcal{F}_{Q}=\left\langle \Delta^{2}\hat{n}_{a}\right\rangle +\left\langle
\Delta^{2}\hat{n}_{b}\right\rangle +2Cov[\hat{n}_{a},\hat{n}_{b}],
\end{equation}
where $\left\langle \Delta^{2}\hat{n}_{i}\right\rangle =\langle\hat{n}_{i}%
^{2}\rangle-\langle\hat{n}_{i}\rangle^{2}$ $(i=a,b)$, and $Cov[\hat{n}%
_{a},\hat{n}_{b}]=\langle\hat{n}_{a}\hat{n}_{b}\rangle-\langle\hat{n}%
_{a}\rangle\langle\hat{n}_{b}\rangle$. $\langle\cdot\rangle$ and
$\langle\Delta^{2}\cdot\rangle$ are the respective average and the variance
calculated in the state $\left\vert \psi_{S}\right\rangle $.

In the presence of photon losses and phase diffusion simultaneously, the QFI
can be obtained from the extend method of Eq. (\ref{CQ}). With Eq.
(\ref{Kraus}), the Hermitian operators $\hat{H}_{1,2}$ become
\begin{align}
\hat{H}_{1}  &  =\sum_{l_{a}=0}^{n_{a}}\sum_{l_{b}=0}^{n_{b}}\frac{d\hat{\Pi
}_{l_{a}l_{b}}^{\dagger}}{d\phi}\frac{d\hat{\Pi}_{l_{a}l_{b}}}{d\phi}=\frac
{1}{4}[(1+\lambda)\nonumber\\
&  \times(\hat{n}_{a}+\hat{n}_{b}-\gamma_{a}^{\prime}\hat{S}_{a}-\gamma
_{b}^{\prime}\hat{S}_{b})+2\lambda(\frac{\hat{p}_{E_{a}}}{4\beta_{a}}%
+\frac{\hat{p}_{E_{b}}}{4\beta_{b}})]^{2}\nonumber\\
&  +\frac{1}{4}(1+\lambda)^{2}[\eta_{a}(\gamma_{a}^{\prime})^{2}\hat{S}%
_{a}+\eta_{b}(\gamma_{b}^{\prime})^{2}\hat{S}_{b}],
\end{align}
and%
\begin{align}
\hat{H}_{2}  &  =i\sum_{l_{a}=0}^{n_{a}}\sum_{l_{b}=0}^{n_{b}}\frac{d\hat{\Pi
}_{l_{a}l_{b}}^{\dagger}}{d\phi}\hat{\Pi}_{l_{a}l_{b}}=-\frac{1}{2}%
[(1+\lambda)\nonumber\\
&  \times(\hat{n}_{a}+\hat{n}_{b}-\gamma_{a}^{\prime}\hat{S}_{a}-\gamma
_{b}^{\prime}\hat{S}_{b})]-\lambda(\frac{\hat{p}_{E_{a}}}{4\beta_{a}}%
+\frac{\hat{p}_{E_{b}}}{4\beta_{b}}),
\end{align}
where%
\begin{equation}
\text{ }\hat{S}_{j}(\eta_{j})=(1-\eta_{j})\hat{n}_{j},\text{ }(j=a\text{, }b).
\end{equation}

Then from Eq.~(\ref{CQ}), the upper bound $\mathcal{C}_{Q}$ of state
($S+E+E^{\prime}$) can be worked out:%
\begin{align}
\mathcal{C}_{Q}  &  =(1+\lambda)^{2}\{\sum_{i=a,b}[(1-\gamma_{i}^{\prime
}(1-\eta_{i}))^{2}\left\langle \Delta^{2}\hat{n}_{i}\right\rangle +\eta
_{i}(\gamma_{i}^{\prime})^{2}\nonumber\\
\times &  (1-\eta_{i})\left\langle \hat{n}_{i}\right\rangle ]+2[1-\gamma
_{a}^{\prime}(1-\eta_{a})][1-\gamma_{b}^{\prime}(1-\eta_{b})]\nonumber\\
&  \times Cov[\hat{n}_{a},\hat{n}_{b}]\}+\frac{\lambda^{2}}{8\beta_{a}^{2}%
}+\frac{\lambda^{2}}{8\beta_{b}^{2}},
\end{align}
where $\gamma_{i}^{\prime}=1+\gamma_{i}$ ($i=a,b$), $\langle\cdot\rangle$ and
$\langle\Delta^{2}\cdot\rangle$ are the respective average and the variance
calculated in the state $\left\vert \psi_{S}\right\rangle \left\vert 0_{E_{a}%
}\right\rangle \left\vert 0_{E_{b}}\right\rangle $. We minimize the above
$\mathcal{C}_{Q}$\ by the parameters $\gamma_{j}^{\prime}$ $(j=a$, $b)$ and
$\lambda$. The minimum of $\mathcal{C}_{Q}$\ is reached for%
\begin{align}
\gamma_{opt,i}^{\prime}  &  =\frac{\left\langle \Delta^{2}\hat{n}%
_{i}\right\rangle ^{\prime}+B_{j}Cov[\hat{n}_{a},\hat{n}_{b}]/(A_{j}+B_{j}%
)}{(1-\eta_{i})\left\langle \Delta^{2}\hat{n}_{i}\right\rangle ^{\prime}%
+\eta_{i}\left\langle \hat{n}_{i}\right\rangle }\text{ }(i,j=a,b\text{
}\nonumber\\
\text{and }i  &  \neq j),\\
\lambda_{opt}  &  =-\frac{8\widetilde{\mathcal{C}}_{Q}\beta_{a}^{2}\beta
_{b}^{2}}{8\widetilde{\mathcal{C}}_{Q}\beta_{a}^{2}\beta_{b}^{2}+\beta_{a}%
^{2}+\beta_{b}^{2}},
\end{align}
where
\[
\widetilde{\mathcal{C}}_{Q}=\sum_{\substack{i,j=a,b\\i\neq j}}\left[
T_{ij}^{2}\left\langle \Delta^{2}\hat{n}_{i}\right\rangle +K_{ij}%
^{2}\left\langle \hat{n}_{i}\right\rangle \right]  +2T_{ab}^{2}T_{ba}%
^{2}Cov[\hat{n}_{a},\hat{n}_{b}],
\]%
\begin{align}
\left\langle \Delta^{2}\hat{n}_{i}\right\rangle ^{\prime}  &  =\left\langle
\Delta^{2}\hat{n}_{i}\right\rangle -\frac{B_{j}}{A_{j}+B_{j}}\frac
{Cov^{2}[\hat{n}_{a},\hat{n}_{b}]}{\left\langle \Delta^{2}\hat{n}%
_{j}\right\rangle },\text{ }(i,j=a,b\text{ }\nonumber\\
\text{and }i  &  \neq j),\nonumber\\
T_{ij}  &  =\frac{A_{i}-\frac{A_{j}B_{i}}{A_{j}+B_{j}}\times\frac{\left\langle
\Delta n_{j}\right\rangle }{\left\langle \Delta n_{i}\right\rangle }J}%
{A_{i}+B_{i}(1-\frac{B_{j}}{A_{j}+B_{j}}J^{2})},\nonumber\\
K_{ij}  &  =\frac{\sqrt{B_{i}}(1+\frac{A_{j}}{A_{j}+B_{j}}\times
\frac{\left\langle \Delta n_{j}\right\rangle }{\left\langle \Delta
n_{i}\right\rangle }J-\frac{B_{j}J^{2}}{A_{j}+B_{j}})}{A_{i}+B_{i}%
(1-\frac{B_{j}}{A_{j}+B_{j}}J^{2})},\nonumber\\
J  &  =\frac{Cov[\hat{n}_{a},\hat{n}_{b}]}{\left\langle \Delta\hat{n}%
_{a}\right\rangle \left\langle \Delta\hat{n}_{b}\right\rangle },\text{ }%
A_{i}=\frac{\left\langle n_{i}\right\rangle }{\left\langle \Delta^{2}%
n_{i}\right\rangle },\text{ }B_{i}=\frac{1-\eta_{i}}{\eta_{i}},\text{\ }%
\nonumber\\
(i  &  =a,b).
\end{align}
where $\widetilde{\mathcal{C}}_{Q}$ is the OFI of the state ($S+E^{\prime}$).

\begin{figure}[tbh]
\centerline{\includegraphics[scale=0.42,angle=0]{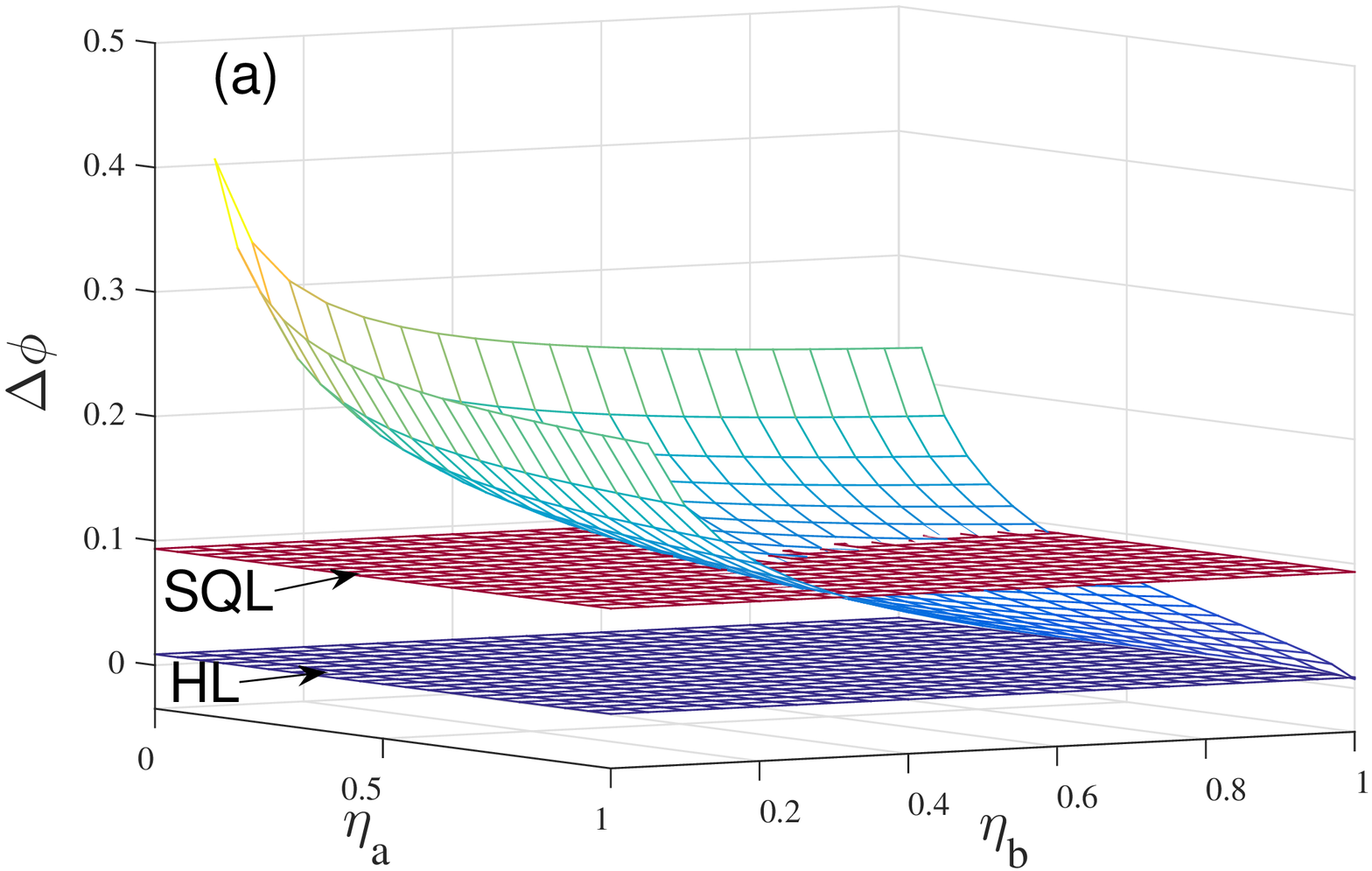}}
\centerline{\includegraphics[scale=0.42,angle=0]{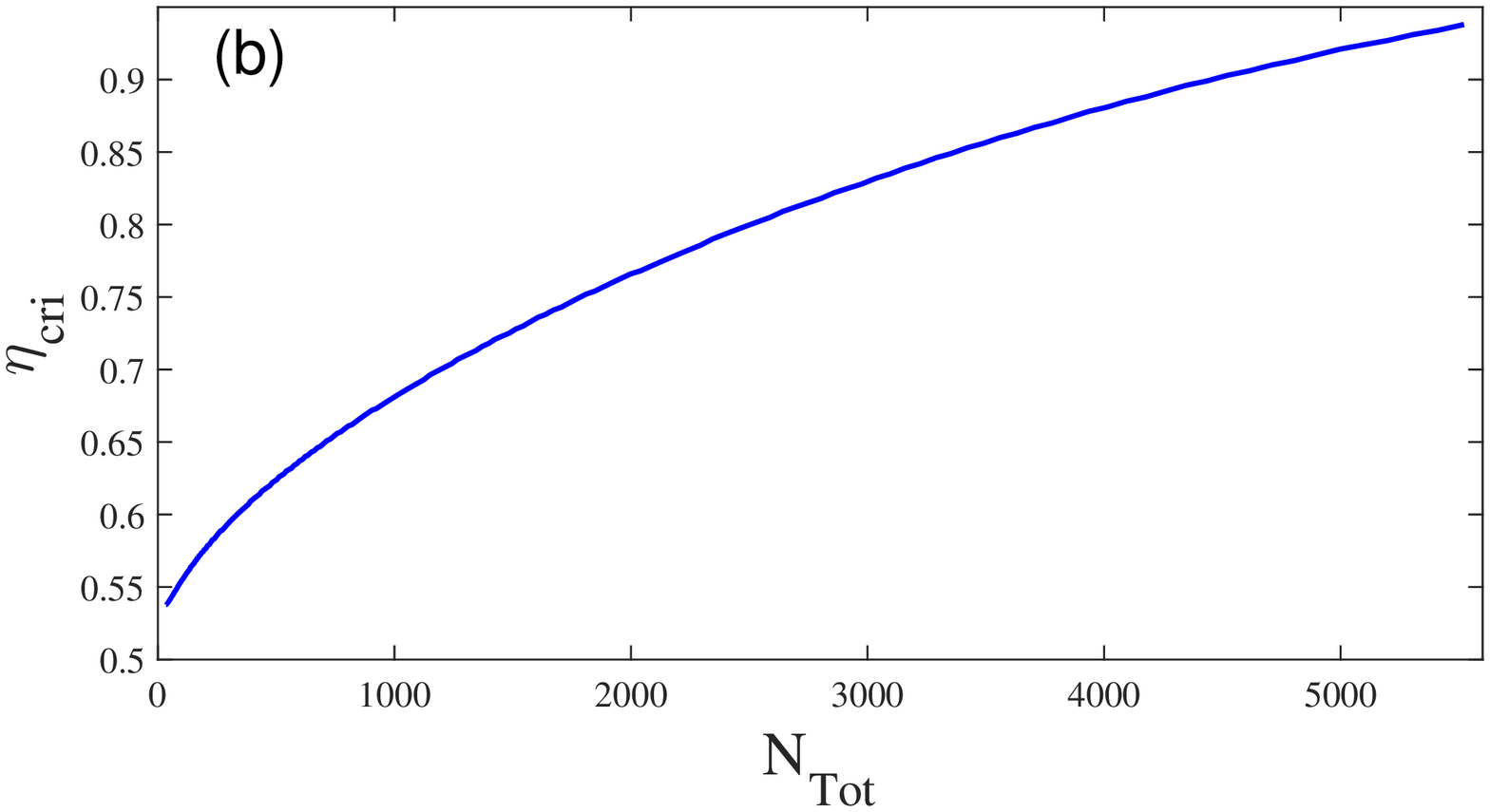}} \caption{(a)
Phase sensitivity $\Delta\phi$ versus photon losses in both arms $\eta_{a}$
and $\eta_{b}$, where $g=2$, $r=1$, $\left\vert \alpha\right\vert ^{2}%
={e^{2r}}/{4}$. (b) The critical $\eta_{\mathrm{cri}}$ as a function of the
total mean photon number inside the interferometer $N_{\mathrm{Tot}}$ where
$g=2$, and $\eta_{a}$=$\eta_{b}$. Given a $N_{\mathrm{Tot}}$, the phase
sensitivity $\Delta\phi$ beat the SQL when $\eta>\eta_{\mathrm{cri}}$.}%
\label{fig4}%
\end{figure}\begin{figure}[tbh]
\centerline{\includegraphics[scale=0.4,angle=0]{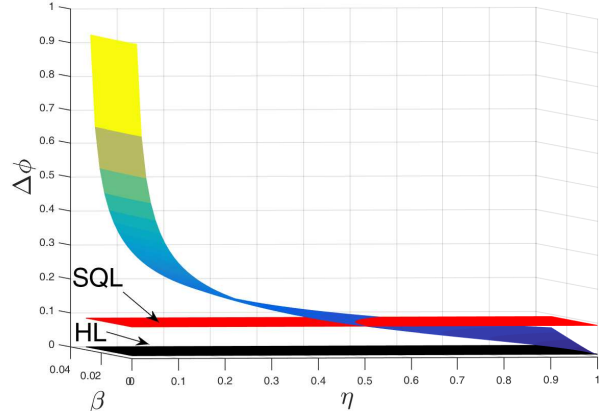}} \caption{Phase
sensitivity $\Delta\phi$ versus photon losses $\eta$ and phase diffusion
$\beta$ in both arms, where $g=2$, $r=1$, $\left\vert \alpha\right\vert
^{2}={e^{2r}}/{4}$. SQL and HL denote the standard quantum limit and
Heisenberg limit, respectively. }%
\label{fig5}%
\end{figure}

Substituting those optimal results $\gamma_{opt,i}^{\prime}$ and
$\lambda_{opt}$ into $\mathcal{C}_{Q}$, the minimal $\mathcal{C}_{Q}$\ is
given by
\begin{equation}
\mathcal{C}_{\phi}=\frac{\widetilde{\mathcal{C}}_{Q}(\beta_{a}^{2}+\beta
_{b}^{2})}{8\widetilde{\mathcal{C}}_{Q}\beta_{a}^{2}\beta_{b}^{2}+\beta
_{a}^{2}+\beta_{b}^{2}},
\end{equation}
and the corresponding phase sensitivity is%
\begin{equation}
\Delta\phi\geq\sqrt{\frac{1}{\widetilde{\mathcal{C}}_{Q}}+\frac{8\beta_{a}%
^{2}\beta_{b}^{2}}{\beta_{a}^{2}+\beta_{b}^{2}}}. \label{phase-total}%
\end{equation}
The phase sensitivity shows the decoherence effect on the phase estimation,
which is dependent on the photon loss coefficients $\eta_{i}$, and the phase
diffusion noise coefficients $\beta_{i}$ ($i=a,b$). When the phase diffusion
noise coefficients $\beta_{a}=\beta_{b}=0$, the above Eq. (\ref{phase-total})
can reduce to the result of only photon loss exists \cite{Gong17}
\begin{equation}
\Delta\phi\geq\sqrt{1/\widetilde{\mathcal{C}}_{Q}}.
\end{equation}
On the other hand, when the photon loss coefficients $\eta_{a}=\eta_{b}=0$,
the result reduces to
\begin{equation}
\Delta\phi\geq\sqrt{1/\mathcal{F}_{Q}+8\beta_{a}^{2}\beta_{b}^{2}/(\beta
_{a}^{2}+\beta_{b}^{2})}.
\end{equation}
It also coincides with bound obtained in Ref.~\cite{Escher12} when $\beta
_{a}=\beta_{b}=\beta$. If $\eta_{i}=0$ and $\beta_{i}=0$ ($i=a,b$), we can
recover the result of lossless case $\Delta\phi\geq\sqrt{1/\mathcal{F}_{Q}}$.

Considering a coherent light ($\alpha=\left\vert \alpha\right\vert
e^{i\theta_{\alpha}}$) combined with a single-mode squeezed vacuum light
($|0,\varsigma\rangle$ with $\varsigma=r\exp(i\theta_{\varsigma})$ is the
squeezing parameter) input, and $g$ describing the strength in NBS process, we
analyze the phase sensitivity numerically. Only considering the effect of
phase diffusion, the phase sensitivity as a function of mean total photon
number $N_{\mathrm{Tot}}$ with different phase diffusion coefficients is shown
in Fig.~\ref{fig3}(a), where the SQL and HL denote the standard quantum limit
and Heisenberg limit, respectively. The optimal sensitivity of phase
estimation could still beat the SQL when $\beta<\beta_{\mathrm{cri}}$ with the
subscript \textquotedblleft cri\textquotedblright\ stands for critical value.
The critical $\beta_{\mathrm{cri}}$ is dependent on the total mean photon
number, and the $\beta_{\mathrm{cri}}$ versus the $N_{\mathrm{Tot}}$ is shown
in Fig.~\ref{fig3}(b). Only considering the effect of photon losses, the phase
sensitivity as a function of photon losses in both arms $\eta_{a}$ and
$\eta_{b}$ is shown in Fig.~\ref{fig4}(a), and the $\eta_{\mathrm{cri}}$
versus the $N_{\mathrm{Tot}}$ in Fig.~\ref{fig4}(b). The SU(1,1)
interferometer can tolerate $10\%$ the photon loss when $N_{\mathrm{Tot}%
}<4500$. The phase sensitivity as a function of photon losses $\eta$
($\eta_{a}=\eta_{b}=\eta)$ and phase diffusion $\beta$\ ($\beta_{a}=\beta
_{b}=\beta$) in both arms is shown in Fig.~\ref{fig5}. It is demonstrated that
the phase precision estimation is more tolerant with loss of light field
compared with phase diffusion.

\section{Conclusions}

In conclusion we have studied the effects of decoherence on the phase
sensitivity in an SU(1,1) interferometer. Based on the generalized phase
transform including phase diffusion coefficients, we derived the phase
sensitivity of an SU(1,1) interferometer in the presence of phase diffusion
and photon losses simultaneously. The loss of light field and phase diffusion
degraded the measure precision, and we have given the critical values where
the phase sensitivity below the SNL in the presence of phase diffusion or
photon losses.

\section{Acknowledgements}

This work is supported by the National Key Research Program of China under
Grant No.~2016YFA0302001 and NSFC Grants No.~11474095, No.~91536114, No.
11574086, No.~11654005, and Natural Science Foundation of Shanghai
No.~17ZR1442800, and the Shanghai Rising-Star Program 16QA1401600.


\end{document}